\documentclass[a4paper,12pt]{article}
\usepackage[english]{babel}
\usepackage{graphics}
\begin{document}

\title{Invisible Higgs Boson Decay into Massive Neutrinos of 4th Generation}

\author{K. Belotsky$^*$, D. Fargion$^{\dag}$, M. Khlopov$^*$$^{\#}$, R. Konoplich$^{\ddag}$
\\and K. Shibaev$^*$}
\maketitle

\noindent
\thanks{$^*$ : Centre for Cosmoparticle physics "Cosmion", Miusskaya Pl. 4,
125047, Moscow, Russia\\
$^{\dag}$: Physics department, Universita' degli studi "La
Sapienza", 5, Piazzale Aldo Moro - I 00185 Roma, Italy; INFN Roma, Istituto
Nazionale di Fisica Nucleare, Italy \\
$^{\#}$: Keldysh Institute of Applied Mathematics, Miusskaya Pl. 4, 125047, Moscow,
Russia\\
$^{\ddag}$: Department of Physics, New York University, New York, NY 10003;
Department of Physics, Manhattan College, Riverdale, New York, NY 10471}

\begin{abstract}Results from several recent experiments provide indirect evidences
in the favor of existence of a 4th generation neutrino. Such a neutrino of 
mass m about 50 GeV is compatible with current physical and astrophysical constraints and well 
motivated in the framework of superstring phenomenology. If sufficiently stable the 
existence of such a neutrino leads to the drastic change of Higgs boson physics: for a 
wide range of Higgs boson masses the dominant mode of Higgs boson decay is 
invisible and the branching ratios for the most promising modes of Higgs boson search 
are significantly reduced. The proper strategy of Higgs boson searches in 
such a framework is discussed. It is shown that in the same framework the absence
of a signal in the search for invisible Higgs boson decay at LEP means either that the 
mass of the Higgs ($M_H$) is greater than 113.5 GeV or that the mass difference $|M_H - 2m|$ 
is small.
\vspace{1pc}
\end{abstract}

The existence of Higgs boson is the necessary consequence of Higgs 
mechanism of electroweak symmetry breaking. Proportionality of its Yukawa coupling 
constants to fermion mass is its important property. Such property is shared by wider 
class of models of electroweak symmetry breaking. For example, it is shared by 
technipion in Technicolor models. So the existence of scalar boson with Yukawa 
coupling constants proportional to fermion masses seems to be a general feature of
electroweak symmetry breaking mechanisms in the standard model. The possibility 
for the existence of new heavy elusive fermions, dominating in Higgs boson decays,
leads to drastic change in the strategy of Higgs boson searches. The idea on the 
dominant invisible modes of Higgs boson decays was discussed in the framework 
of Majoron, SUSY and low scale gravity models in \cite{invisibleH} (and references therein),
however the simplest 
possibility is the existence of massive neutrino of 4th generation. The measured 
Z-boson width excludes the existence of 4th neutrino with the mass below 45 GeV.
However, the detailed analysis of 4th generation effects in the standard model 
parameters opens the window for its existence, provided that the mass of 4th neutrino 
is below around W-boson mass \cite{Vysotsky}. A fit \cite{Vysotsky} of the precision electroweak 
data is compatible with the 4th generation neutrino mass $\ m\sim $ 50 GeV.
However, the allowed range for 4th neutrino masses may be wider,
if the 4th neutrino is not accompanied by 4th generation quarks (as it can take place in
Dp-brane phenomenology, naturally excluding quarks of 4th generation but leaving the 
room for the 4th generation of leptons \cite{Aldazabal}). It even can not be accompanied 
by a charged lepton, as it is assumed in some models of neutrino mass \cite{Hung}. 
Provided that it is sufficiently metastable, or it has only invisible decay modes, 
4th neutrino with the mass around 50 GeV could escape detection by products 
of its decay in LEP. The possibility \cite{5} to analyze the LEP data on single gamma
events, corresponding to a 4th neutrino pair production in the reaction 
$e^{+}e^{-}\rightarrow N\bar{N}\gamma $ for $m>50$ GeV is still not realized.

If 4th neutrino is sufficiently long-living or even absolutely stable 
its primordial gas from the early Universe can survive to the present time and 
concentrate in the Galaxy \cite{5}, \cite{6}. It was shown \cite{7} that galactic fluxes of 
4th neutrino can lead to the effect of indirect WIMP searches compatible with the 
DAMA data, and the effects of 4th neutrino-antineutrino annihilation in the Galaxy can 
explain \cite{8}, \cite{9} the galactic gamma background with energies above 1 GeV, 
observed by EGRET. The latter possibility is strengthened, provided that the 
4th generation quarks and leptons possess the new strictly conserved gauge charge, 
as it can naturally follow from heterotic string phenomenology \cite{10}. In that case 
4th neutrino annihilation in Galaxy can explain the positron anomaly in electron 
component of cosmic rays \cite{10}, \cite{11}. It was shown \cite{12} that the capture 
of 4th neutrinos by Earth can lead to the underground neutrino flux, accessible, in 
principle, to underground neutrino detectors. 

In the present note we draw attention to the important role, 
4th neutrino can play in the physics of Higgs boson. The probability 
of Higgs boson decay into $N \bar{N}$ pair is given by
\begin{equation}
\Gamma(H \to N\bar{N})=\frac{\sqrt{2}}{8\pi}Gm^2M_H \left( 1-\frac{4m^2}{M_H^2} 
\right)^{3/2},
\end{equation}
where $G$ is the Fermi constant, $m$ and $M_H$ are the masses of the neutrino 
and Higgs boson respectively. This mode should be compared with the probabilities of 
the most important $b\bar{b}$ and $W W$ modes of Higgs boson decay.

In Fig.1 the dependence on Higgs boson mass 
for branching ratios for these decay modes and other most contributing modes is given 
for two values of the neutrino mass, $m=50$ GeV (Fig 1a) and $m=70$ GeV (Fig 1b). 
One easily finds, that the dominance of $N\bar{N}$ mode in the Higgs boson
width for Higgs boson masses up to 160 GeV naturally follows from the fact
that the mass of $N$ is by the order of magnitude larger than the mass of
b-quark also taking into account the number of colored quark states in the
b-mode (3), a reduction factor $\approx 2$ due to QCD corrections \cite{QCD} and the
phase volume difference for $N\bar{N}$ and $b\bar{b}$ channels. It leads to 
the branching ratio of $N\bar{N}$ channel between 90 and 95 per cent in the total Higgs 
boson width, if the mass of Higgs boson is below the threshold for $WW$ mode. 
However, even at higher masses of Higgs boson the 4th neutrino channel is significant. 
Note that if the masses of 4th lepton and 4th generation quarks are near the existing 
lower limits, the respective decay modes would be important for the Higgs boson mass 
above 260 GeV. In the considered Higgs boson mass range the 4th generation fermions affect
Higgs decay modes $H\to \rm{gg},\,\gamma\gamma$ by loop diagrams. Here their probabilities
were estimated assuming 4th charge lepton mass $m_E=100$ GeV, 4th up- and down- 
quark masses $m_U=m_D=130$ GeV. It takes into account the difference in probabilities of 
these decays in comparison with previous analogous estimates for three fermion generations.

\begin{figure}
\center{\includegraphics*[22cm, 8cm]{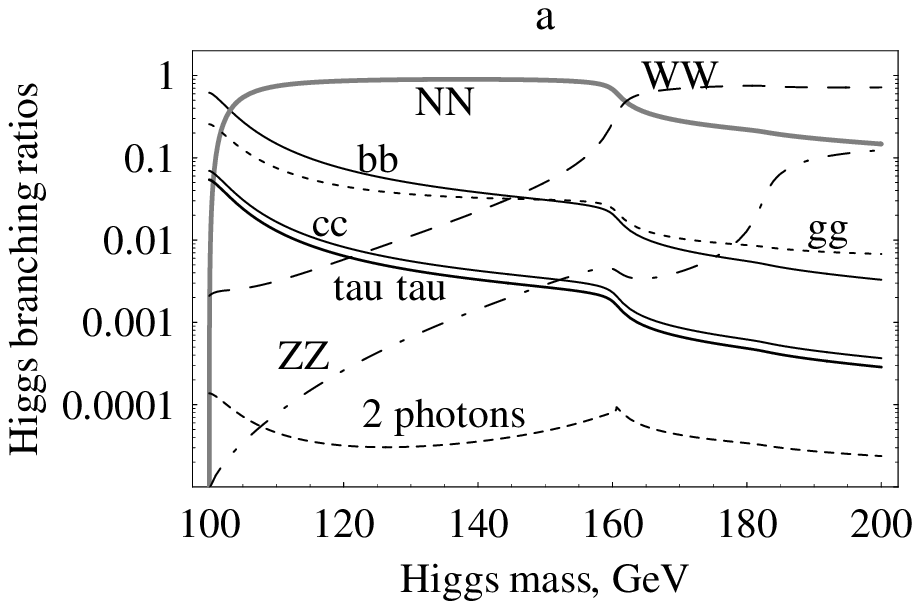}}
\center{\includegraphics*[22cm, 8cm]{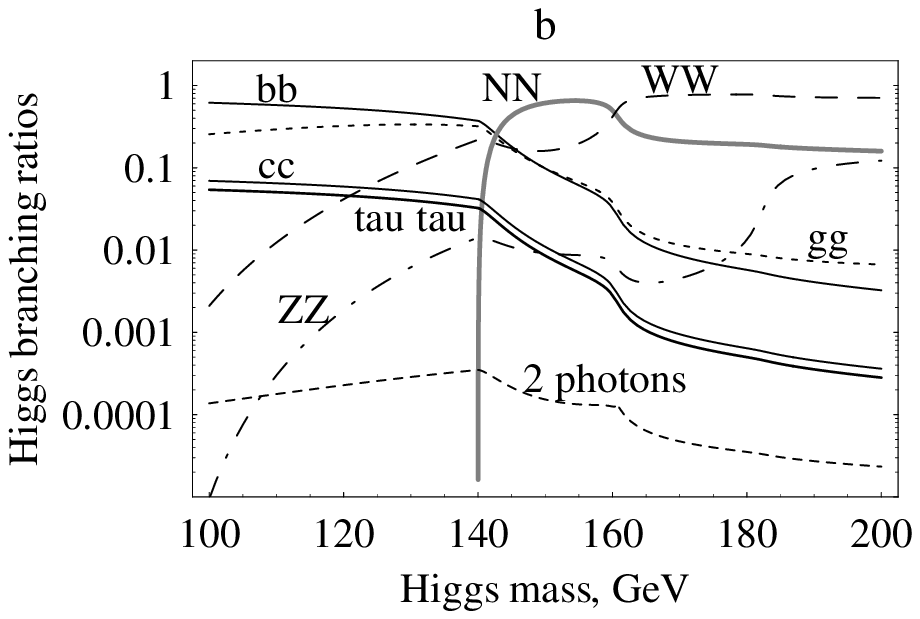}}
\caption{Branching ratios of the Higgs decay modes: $H\to N\bar{N},\, b\bar{b},\, WW,\, ZZ,\,
c\bar{c},\,\tau^+\tau^-,$ gg, $\gamma\gamma$. $a)$ $m=50$ GeV, $b)$ $m=70$ GeV.}
\end{figure}

In the case, when the $N\bar{N}$ decay mode dominates, Higgs boson search can be 
effectively undertaken in the search for acoplanar lepton pairs or acoplanar jets, 
arising from the Higgs-struhlung reaction \cite{7},\cite{8}
\begin{equation}
e^+e^- \to ZH
\end{equation}
with successive Z-boson decay into charged lepton pair, and elusive 
Higgs boson decay into the pair of $N\bar{N}$. The total cross section for this reaction 
is given by
\begin{equation}
\sigma (e^+e^- \to ZH)=\frac{(g_V^2+g_A^2)G^2M_Z^6}
{6\pi\sqrt{s}}\frac{|\overrightarrow{q_1}|}{(s-M_Z^2)^2+M_Z^2\Gamma_Z^2}
\left(1+\frac{1}{2}\frac{q_{10}^2}{M_Z^2}\right),
\end{equation}
where $g_V=-1+4 \sin ^2 \theta_W $, $g_A=-1$, $q_{10}=(s+M_Z^2-M_H^2)/
(2\sqrt{s})$, $s$ being squared center-mass energy, $|\overrightarrow{q_1}|=\sqrt{q_{10}^2-
M_Z^2}$. The differential cross section for charged leptons from successive 
Z-boson decay is given by
\begin{eqnarray}
\frac{l_0d\sigma _{\pm }}{d^3l} &=&\frac 3{2^5\pi ^2}\beta _N 
\beta _l \frac{M_Z^4 G^2 D_Z}{C_V}\frac 1{\sqrt{s}l_0}\{2l_0(q_{10}-l_0) 
C_V^2\mp C_A^2M_Z^2 cos\theta + \nonumber 
\\&&+\cos^2 \theta C_V^2[M_Z^2-2l_0(q_{10}-l_0)]\},
\end{eqnarray}
where $C_V=g_V^2+g_A^2$, $C_A=2g_V g_A$, $D_Z=\frac{1}{(s-M_Z^2)^2+M_Z^2
\Gamma_Z^2}$, $\beta _N$ and $\beta _l$ are the $H\to NN$ and $Z\to l^+l^-$ 
 branching ratios respectively, $\theta$ is the angle between momenta of 
the initial electron and 
the final negative lepton, the kinematic limits of the considered process are 
$M_Z+M_H\leq \sqrt{s}$, $\frac{1}{2}[q_{10}-\left| \overrightarrow{q_1}\right| ]\leq 
l_0\leq \frac{1}{2}[q_{10}+\left| \overrightarrow{q_1}\right| ]$. The above formulae 
can be used for the case of visible modes of Z-boson decay to $\mu \bar{\mu}$ 
and $\tau \bar{\tau}$, as well as for 2-jet events from quark antiquark channels 
of Z-boson decay (replacing $C_V$, $C_A$ by appropriate values and taking into account 
3 color degrees of freedom). 
In the case of electron-positron mode of Z-boson decay an interference 
 diagram should be taken into account. 

Fig.2 shows the total cross sections of $\mu ^{+}\mu ^{-}$ pair production
for the Higgs-struhlung reaction (Eq.2) with the Higgs boson decaying
invisibly into $N\overline{N}$ of mass 50 GeV. One can see that the total
cross section is within the range of LEP collider. The LEP working group
presented results of a search for a Higgs boson, produced at the Standard
Model rate, decaying into invisible particles \cite{13},\cite{14}. No statistically
significant excess has been observed when compared to a background
prediction. Assuming that the Higgs boson decays only into invisible states
a lower bound has been set on its mass at 95\% CL of 114.4 GeV \cite{13}. In
the case of 4th generation neutrino of $m=50$ GeV taking into account the
branching ratios one can infer from the LEP data \cite{13} the lower limit
on Higgs mass  $\approx $ 113.5 GeV.  If a 4th generation neutrino of mass $%
m\approx 50$ GeV exists then the absence of the signal at LEP\ means either that 
the mass of the Higgs $M_H>113.5$ GeV or that the mass difference $\left\vert
2m-M_{H}\right\vert $ is small leading to a significant phase space suppression of
the invisible Higgs decays. 

\begin{figure}
\center{\includegraphics*[22cm, 8cm]{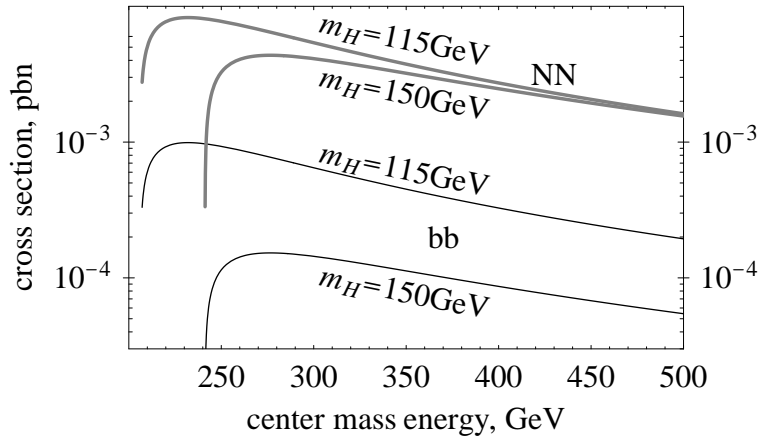}}
\caption{Total cross sections of the processes $e^+e^- \to ZH\to l^+l^-\,b\bar{b},\,
l^+l^-\,N\bar{N}$ in dependence on center mass energy of colliding $e^+\,e^-$ for 
$M_H=115$ GeV and $M_H=150$ GeV when $m=50$ GeV.}
\end{figure}

Actually, the predicted effect of elusive Higgs 
boson assumes the lifetime of 4th neutrino to be $>10^{-7}$ sec. If this lifetime
exceeds the age of the Universe, astrophysical 
search for 4th neutrino effects will be of special interest. It turns out that the inverse 
effect of Higgs boson on the predicted astrophysical effects of 4th neutrinos is 
restricted by the very narrow interval of 4th neutrino and Higgs mass ratio \cite{15}. 
The mass difference between $2m$ and $M_{H}$ should be negative and less 
than 3-4 GeV to influence significantly the primordial $N\bar{N}$ concentration owing 
to the Higgs boson channel of their annihilation in the period of their freezing out in 
early Universe. So, for the mass of neutrino 50 GeV and the Higgs boson mass 
114 GeV the role of Higgs boson is elusive in the calculations of 4th neutrino 
freezing out in the early Universe, as well as in the effects of 4th neutrino annihilation 
in the Galaxy. The detailed discussion of Higgs boson effect on the astrophysical 
signatures of 4th neutrinos will be given in a separate paper. 

To conclude, the existence of scalar boson with Yukava coupling proportional to 
fermion mass is the important signature for the mechanism of electroweak 
symmetry breaking. The existence of massive 4th neutrino makes elusive the dominant 
mode of decay of this boson, and the strategy of Higgs boson search should take 
into account this possibility. In particular, the
presented results mean that for LHC in the mass region $M_{H}$ $\simeq
115-160$ GeV the gluon fusion process $\rm{gg}\to H$ is not dominating and one has
to search for lepton or jet pair + missing energy from reactions $qq\to qqH,\,WH,\,ZH$. 
The positive result of such search will not only prove one of the corner stones
of the standard model, but it will also prove the existence of physics beyond the 
standard model, as well as it will make the hypothesis of 4th neutrino deserving serious
attention. The experimental proof that the ratio of the elusive mode to b-quark one of Higgs
boson decay is as it is predicted for 4th neutrino will strongly favor this hypothesis as 
compared with other possible models for invisible Higgs boson. The set of astrophysical
signatures provides the complete test of the hypothesis on massive stable 4th neutrino.

The work was partially performed in the framework of Russian State contract 
$40.022.1.1.1106$, with partial support of Cosmion-ETHZ, AMS-Epcos collaborations, 
of grant $00-15-96699$ of support for Russian scientific schools, RFBR grant 
$02-02-17490$ and of grant of Russian Universities. One of us (M.Yu.Kh) expresses 
his gratitude to Rome University ``La Sapienza'' and to IHES for hospitality.

\end{document}